\documentclass[12pt]{article}

\newcommand{\be}{\begin{equation}}
\newcommand{\ee}{\end{equation}}
\newcommand{\bea}{\begin{eqnarray}}
\newcommand{\eea}{\end{eqnarray}}
\newcommand{\bean}{\begin{eqnarray*}}
\newcommand{\eean}{\end{eqnarray*}}
\newcommand{\brray}{\begin{array}}
\newcommand{\erray}{\end{array}}
\newcommand{\ben}{\begin{equation}{nonumber}}
\newcommand{\een}{\end{equation}{nonumber}}

\newtheorem{dfn}{Definition}[section]
\newtheorem{thm}[dfn]{Theorem}
\newtheorem{lmma}[dfn]{Lemma}
\newtheorem{ppsn}[dfn]{Proposition}
\newtheorem{crlre}[dfn]{Corollary}
\newtheorem{xmpl}[dfn]{Example}
\newtheorem{rmrk}[dfn]{Remark}

\newcommand{\bdfn}{\begin{dfn}}
\newcommand{\bthm}{\begin{thm}}
\newcommand{\blmma}{\begin{lmma}}
\newcommand{\bppsn}{\begin{ppsn}}
\newcommand{\bcrlre}{\begin{crlre}}
\newcommand{\bxmpl}{\begin{xmpl}}
\newcommand{\brmrk}{\begin{rmrk}}

\newcommand{\edfn}{\end{dfn}}
\newcommand{\ethm}{\end{thm}}
\newcommand{\elmma}{\end{lmma}}
\newcommand{\eppsn}{\end{ppsn}}
\newcommand{\ecrlre}{\end{crlre}}
\newcommand{\exmpl}{\end{xmpl}}
\newcommand{\ermrk}{\end{rmrk}}

\newcommand{\al}{\alpha}

\newcommand{\cla}{{\cal A}}
\newcommand{\clb}{{\cal B}}

\newcommand{\cld}{{\cal D}}
\newcommand{\cle}{{\cal E}}
\newcommand{\clf}{{\cal F}}

\newcommand{\clk}{{\cal K}}
\newcommand{\cll}{{\cal L}}

\newcommand{\clv}{{\cal V}}

\newcommand{\clz}{{\cal Z}}

\def\a*{{\cal A}_{h,*}}
\def\B{{\cal B}(h)}
\def\B1{{\cal B}_1(h)}
\def\b{{\cal B}^{s. a. }(h)}
\def\b1{{\cal B}^{s. a. }_1(h)}

\newcommand{\ot}{\otimes}

\newcommand{\raro}{\rightarrow}

\begin{document}
\begin{center}
{\large {\bf A Remark on the structure  of symmetric quantum
dynamical semigroups on von Neumann
 algebras}}\\
 \vspace{6ex}
\begin{center}
Sergio Albeverio \footnote{ SFB 256; SFB 237; BiBoS; CERFIM(Locarno); Acc. Arch.; USI(Mendriso)}and
  Debashish Goswami  \footnote{ Alexander von Humboldt Fellow}\\

 \end{center}
\vskip 1 true cm
  \end{center}

\begin{abstract}
We study the structure of the generator of a symmetric, conservative  quantum dynamical semigroup
 with norm-bounded generator on a von Neumann algebra equipped
 with a faithful semifinite trace.  For von Neumann algebras with abelian commutant (i.e. type I von Neumann
  algebras), we give a necessary and sufficient algebraic condition for the generator of such a semigroup to be
   written as a sum of square of self-adjoint derivations of the von Neumann algebra.   This generalizes some of the results obtained
      by Albeverio, H$\phi$egh-Krohn and Olsen (\cite{Alb}) for the special case of the finite dimensional matrix
      algebras. We also study similar questions for a class of quantum dynamical semigroups with
       unbounded generators.
\end{abstract}

\section{Introduction}
In \cite{Alb}, among other things trace-symmetric conservative quantum dynamical semigroups
 on the algebra $M_n$ of $n \times n$ matrices are studied.  Our purpose
   in the present note is to study similar questions for a more general (possibly infinite
    dimensional) class of von Neumann algebras, equipped with some faithful semifinite trace.
      We give a complete characterization of the structure of
      the generator of a symmetric (w.r.t. the above-mentioned trace) conservative quantum
       dynamical semigroup with bounded generator, and also obtain some results for certain
        semigroups with unbounded generators.

           First of all, we shall recast some of the results obtained in \cite{Alb} in a slightly
            new language. We recall that the basic fact used in section 2 of \cite{Alb} is the
             following structure-theorem for the generator $\cll$ of a trace-symmetric conservative
              quantum dynamical semigroup $T_t$ on $M_n$ :\\
              $$ \cll(a)=\frac{1}{2}\sum_{j=1}^k [\beta_j,[\beta_j,a]]; a \in M_n; $$
               where $k \geq 1$ is some integer and $\beta_j \in M_n$ with $\beta_j^*=-\beta_j;
                tr(\beta_j)=0$. The above result is a simple corollary of the results obtained by Lindblad
                 \cite{Lin}. Let  the self-adjoint derivation $[\beta_j,.]:
                  M_n \raro M_n$ be denoted by $\delta_j$. Then we can write the above as
                  $\cll=\frac{1}{2}\sum_j \delta_j^2$. We shall study in the next section necessary and
                   sufficient conditions under which the generator of a quantum dynamical semigroup
                    (q.d.s. henceforth)on a von Neumann algebra can be written in the above form (with possibly infinitely
                     many derivations, where the sum should converge in an appropriate topology). In particular,
                      we shall show that for a general von Neumann algebra with a faithful semifinite trace,
                       symmetry and conservativity alone do not suffice for such a structure, but one
                        needs additional conditions. Then, finally, we shall extend our discussion to
                         some special classes of semigroups with unbounded generators too.

                               \section{Structure and dilation of symmetric conservative q.d.s. on more
                                general von Neumann algebras : bounded generator case}

                       Let $\cla \subseteq \clb(h)$ (where $h$ is a separable Hilbert space)
                        be a von Neumann algebra with a normal, faithful, semifinite
                        trace $\tau$ on it. Let $L^p, 1 \leq p \leq \infty $ denote the well-known
                         $L^p$-spaces with respect to the trace $\tau$ (cf \cite{Se}, \cite {Nel}).
                           Let $T_t$ be a conservative q.d.s. on $\cla$, with norm-bounded generator $\cll$.
                            We also assume   that  $T_t$ is
                         symmetric w.r.t. $\tau$, in the sense that $\tau(T_t(a)b)=\tau(aT_t(b)) $ for all   $a,b \in
                          \cla$ with $a,b \geq 0$.        By the standard interpolation principle,
                            $T_t$ can be uniquely extended as a continuous map from $L^p$ to $L^p$ for each $p$. We shall
                             not notationally distinguish among these maps, denoting all of them by the same symbol
                              $T_t$ (see, e.g. \cite{CS} and the references
                              therein for more discussion on this).

                              It is clear that the generator $\cll$ of the above q.d.s. will also satisfy the symmetry condition
                               $\tau(\cll(a)b)=\tau(a\cll(b))$ for all $a,b \in L^1 \bigcap L^\infty$, and then by using the duality
                                $(L^1)^* \cong L^\infty$, it is easy to see that $\cll$ extends as
                                 a continuous linear map from $L^1$ to $L^1$, and $\tau(\cll(a)b)=\tau(a\cll(b))$
                                  holds for all $a \in L^1$, $b\in L^\infty$.

                                    Now, it is well-known (\cite{CE}) that $\cll$ can be written as $\cll(a)=\psi(a)
                                    +ka+ak+i[H,a]$, where $\psi : \cla \raro \cla$ is a normal completely positive (CP) map,
                                     $k, H$ are self-adjoint elements of $\cla$. Since it is clear that $\tau((ak+ka+i[H,a])b)
                                     =\tau(a(kb+bk-i[H,b]))$ for all
                                     $a \in L^1$, $b \in L^\infty$,
                                       we conclude that $\psi$ satisfies
                                       $\tau(\psi(a)b)=\tau(a\psi(b))
                               +2i\tau(a[H,b])$. In particular, $\tau(\psi(a))=\tau(a\psi(1))
                                \forall a \in L^1.$

                                     We shall now prove the main result of this section.
                                     \bthm
                                     \label{1}
                                     I.  Let $ H_j; j=1,2,...$ be elements of $\cla$ such that
                                     $H_j^*=-H_j$,  and assume that $\sum_{j=1}^\infty H_j^2$
                        weakly converges. Then there is a normal norm-bounded map $\cll$ from $\cla$
                         to itself given by
                         $$
                        (1) \hspace{8mm}             \cll(a)=\left( \frac{1}{2}\sum_j [H_j,[H_j,a]]\right);
                         $$
                (where the  series in the right hand side above converges weakly) such that
                $T_t=e^{t \cll}$ is a symmetric conservative q.d.s. on $\cla$. Furthermore,
            the following algebraic relation is satisfied :\\
            $$
            (2) \hspace{8mm}            \partial(az,az)=z^*\partial(a,a)z \forall a \in \cla, z \in \clz;
            $$
               where $\partial(x,y) :=\cll(x^*y)-\cll(x^*)y-x^*\cll(y)$ and $\clz$ denotes the centre
                of $\cla. $\\
                (II) (partial converse)\\
                Under the additional assumption that $\cla$ is of type I, the
                 converse of (I) holds, i.e. given any symmetric conservative q.d.s. $T_t$
                 with bounded generator $\cll$ on $\cla$ satisfying the algebraic relation
                  (2), there exist $H_j,j=1,2,...$ satisfying the conditions
                  of (I) and such that (1) is satisfied.

                                    \ethm

{\it Proof :-}\\
(I) The proof of this part is more or less standard and can be found in the literature,
 e.g., \cite{DL}. We, however, give a proof here for the sake of completeness. Let $H_0=\sum_j H_j^2$, which is clearly a self-adjoint element of $\cla$. To show that
$\sum _j [H_j,[H_j,a]]$ weakly converges for all $a \in \cla$, it suffices to prove it for all nonnegative
 elements $a$. Fix $a \in \cla_+$, and define $b_n=-\sum_{j=1}^nH_jaH_j=\sum_{j=1}^nH_j^*aH_j$.
  Clearly, $0 \leq b_n \uparrow$, and $b_n \leq \|a\| (-\sum_{j=1}^\infty H_j^2)$, thus $\sup_n \|b_n\| < \infty$.
  So, $b_n$ must be weakly convergent. Since $\sum_{j=1}^n [H_j,[H_j,a]]=-2\sum_{j=1}^n
   H_jaH_j+ (\sum_{j=1}^n H_j^2)a+a(\sum_{j=1}^nH_j^2)$, it follows that
$\sum _j [H_j,[H_j,a]]$ is weakly convergent. Furthermore, let $\clk$ be any separable Hilbert space with an orthonormal
 basis $\{ e_j,j=1,2,... \}$, and $R : h \raro h \ot \clk$ be a bounded linear map defined by
  $Ru:=\sum_j H_ju \otimes e_j$ (that $R$ is well-defined and bounded follows easily from
   the fact that $\sum_j H_j^2$ is weakly convergent, and it is easy to see that $\|R\|^2=\|\sum_j
   H_j^2\|$). It is then easily verified that $\cll(a)=R^*(a \ot 1_\clk)R-\frac{1}{2}R^*Ra-a\frac{1}{2}R^*R
   $, from which we immediately conclude that $\cll$ is a bounded normal map and $e^{t \cll}$
    is a conservative q.d.s. It remains to prove that $T_t$ is symmetric. First we note that since
    $\tau$ is normal, and for $a,b \in L^1 \bigcap \cla_+$, $0\leq
    =-\sum_{j=1}^n \sqrt{b}H_jaH_j\sqrt{b} \uparrow -\sum_{j=1}^\infty \sqrt{b}H_jaH_j\sqrt{b}$,
     it follows that $\tau(\sum_{j=1}^\infty H_jaH_jb)=
    \tau(\sum_{j=1}^\infty \sqrt{b}H_jaH_j\sqrt{b}) =\sum_{j=1}^\infty
      \tau(\sqrt{b}H_jaH_j\sqrt{b})=\sum_{j=1}^\infty \tau(aH_jbH_j)
      =\tau(a\sum_{j=1}^\infty H_jbH_j)$. The same equality
     will clearly hold for all $a,b \in L^1 \bigcap L^\infty$, and from this a straightforward
      calculation enables us to verify that $\cll$ is symmetric. The symmetry of $T_t$ then
       follows from the fact that $T_t(a)$ is given by the norm-convergent series $\sum_{n=0}^\infty
        \frac{t^n}{n!}\cll^n(a)$. Finally, the relation (2) is verified by a direct
         calculation using the form of $\cll$ given by (1). This completes the proof
          of (I).

          (II.) To prove the converse, we shall proceed  as follows. Since $\cla$ is of type I,
           without loss of generality we can choose $h$ such that the commutant of $\cla$ in $\clb(h)$
            is abelian. Now,  we shall very
           briefly recall the arguments given in \cite{GLSW}, \cite{G} to show that there exist
            a separable Hilbert space $\clk_1$ and a bounded linear map $R$ from
             $h$ to $\clk_1$ such that
              $\cll(a)=R^*(a \ot 1_{\clk_1})R-\frac{1}{2}R^*Ra-a\frac{1}{2}R^*R
       +i[H,a]$, where $H \in \cla$ is self-adjoint, and furthermore $R_i \in \cla \forall i,$
        where $R_i$ are defined by $\langle R_iu,v\rangle=\langle Ru, v\ot e_i\rangle $
        ($\{ e_i \}$ is some orthonormal basis of $\clk_1$). By the well-known result of
        Christensen and Evans (\cite{CE}), we can choose a separable Hilbert space
         $\clk_0$ and $S \in \clb(h,\clk_0)$, a normal $\ast$-homomorphism $\rho: \cla
         \raro \clb(\clk_0)$ and $H \in \cla_{s.a.}$ such that
         $\cll(a)=S^*\rho(a)S-\frac{1}{2}S^*Sa-a\frac{1}{2}S^*S
       +i[H,a]$, with the minimality condition that the linear span of $\{ \al(x)u,x \in \cla,
        u \in h \}$ (where $\al(x)=Sx-\rho(x)S$) is dense in $\clk_0$. Then, we construct
         (c.f., e.g., \cite{G}, \cite{GS}) a normal $\ast$-homomorphism $\rho^\prime$ of the
          commutant $\cla^\prime$ of $\cla$ in $\clk_0$ which satisfies $\rho^\prime(a)\al(x)u
          =\al(x)au \forall a \in \cla^\prime, x \in \cla, u \in h$. It is easy to verify by using
           the algebraic relation (2) that $\rho^\prime$ and $\rho$ agree on the centre $\clz$.
             We note that
                 as  $\cla^\prime$ is abelian,  $\clz=\cla^\prime$.  We
                  choose  a separable Hilbert space $\clk_1$ such that $\rho(x)=\Sigma_1^*(x \ot 1_{\clk_1})\Sigma_1$,
                   where $\Sigma_1 \in \clb(\clk_0, h \ot \clk_1)$
                   is an isometry (such a choice is possible by
                   the general result on the structure of a normal
                    $\ast$-homomorphism of a von Neumann algebra),
                                        and let $R=\Sigma_1 S$. Then  by some simple arguments
                                        as in \cite{GLSW}, \cite{G} we show that the required conditions hold.

                   Now we shall use the assumption of symmetry. We have earlier noted that
                    $\tau(\cll(a)b)=\tau(a \cll(b)) \forall a \in L^1, b \in L^\infty$.
            Let us denote by $\psi : \cla \raro \cla$ the CP map given by $\psi(a)=
             R^* (a \ot 1)R=\sum_{i =1}^\infty R_i^*aR_i$, which is a weakly convergent sum.
             We claim that $\sum_i R_iR_i^*$ also weakly converges and $\psi(a)=\sum_j R_j^*aR_j
             =\sum_j R_jaR_j^*-2i[H,a]$. It is clear from the discussion preceding the present theorem that
              $\tau(\psi(a)b)=\tau(a\{ \psi(b)+2i[H,b]\}) \forall a \in L^1 \bigcap L^\infty, b \in L^\infty$.
              Let $c_n=\sum_{i=1}^n R_iR_i^*$. For any $a \in L^1 \bigcap L^\infty$ with $a \geq 0$,
                we have that
               $\tau(ac_n)=\sum_{i=1}^n  \tau(aR_iR_i^*)=\sum_{i=1}^n \tau(R_i^*aR_i)\leq \tau(\psi(a))
               =\tau(a \psi(1))$. Since any element in $L^1\bigcap L^\infty$ can be canonically decomposed into
                 a linear combination of nonnegative elements of $L^1 \bigcap L^\infty$, by a standard argument we
                 show that $| \tau(ac_n)| \leq C \|a\|_1 \|\psi(1)\|_\infty$ for all $a \in L^1 \bigcap L^\infty$
                      (where $C$ is some numerical constant independent of $a,c_n$). Since
                      $L^1 \bigcap L^\infty$ is dense in $L^1$ in $L^1$-norm, and since
                       $\| c_n \|_\infty =\sup_{a \in L^1, \|a\|_1 \leq 1} |\tau(ac_n)|$,
                        we conclude that $\sup_n \|c_n\|_\infty < \infty$, and  $0 \leq c_n \uparrow,
                        $ which proves that $c_n$ is weakly convergent. Furthermore, by an exactly similar
                         argument we can prove that $\sum_i R_i b R_i^*$ is weakly convergent for every
                          $b \in L^\infty$. Now, by normality of $\tau$, it is easy to see that $\tau(\sum_i
                           aR_ibR_i^*)=\sum_i \tau(aR_ibR_i^*)$ for all $a,b \in L^1\bigcap L^\infty,a,b \geq 0$, and hence the same equality will
                            hold for all $a \in L^1,b \in \cla$. Using this equality  we verify that $\tau(a\psi(b))=
                            \tau(a \{ \sum_j R_jbR_j^*-2i[H,b]\})$ for all $a \in L^1 , b \in \cla$, and hence $\psi(b)=
                             \sum_j R_jbR_j^*-2i[H,b]$, which completes the proof of the claim. Finally, we choose
                              $H_j,j=1,2,...$ as follows :\\
                              $$  H_{2j}=\frac{i}{2}(R_j+R_j^*), H_{2j-1}=\frac{1}{2}(R_j-R_j^*).$$
                              Then  a straightforward computation using the fact that $\sum_j
                               R_jxR_j^*=\sum_j R_j^*xR_j+2i[H,x]$ enables us to conclude that
                                $\cll(x)=\sum_{j=1}^\infty \frac{1}{2}[H_j,[H_j,x]]+i[H^\prime,x] \forall x \in \cla,$
 for some s.a. $H^\prime$. Using again the symmetry of $\cll$, it follows that  $\tau([H^\prime,x]y)=\tau(x[H^\prime,y])
= -\tau([H^\prime,x]y)$ $\forall x,y \in L^1 \bigcap L^\infty$, so that $[H^\prime,x]=0 \forall x \in \cla$. This completes the proof.

                            \brmrk
                            It is shown in \cite{GLSW} that the condition (2) is satisfied by any q.d.s. with norm continuous
                             generator in case $\cla$ is a type I factor. On the other hand, for a function algebra,
                              i.e. $\cla=L^\infty(\Omega,\mu)$ for some measure space $(\Omega,\mu)$, the condition (2)
                               can never be satisfied by any nontrivial q.d.s. with norm continuous generator. Thus,
                                for abelian $\cla$, we conclude that symmetry does not imply the structure given by (1),
                                  since there are plenty of examples of symmetric
                                 nontrivial q.d.s.  with bounded generator on such an algebra. However, if we consider
                                  q.d.s. with unbounded generator on nice function algebras, the above condition is nothing
                                   but locality of the generator (cf \cite{GLSW}).
                          \ermrk

                         \brmrk
                            Under the assumption that $\cla$ is type I, a q.d.s. $T_t$ on
                             $\cla$ with bounded generator $\cll$ admits an Evans-Hudson (EH) dilation with only classical Brownian
                              motion as noise, in the sense that   the EH dilation $j_t$ satisfies
                               a q.s.d.e. of the form $dj_t(x)=j_t(\cll(x))dt+
                                \sum_{i=1}^\infty j_t(\theta_i(x))(da_i(t)+da_i^\dagger(t)); j_0(x)=x$, where $\theta_i : \cla
                                 \raro \cla$ are bounded linear maps, if and only if
                                 the algebraic relation (2) holds and $T_t$ is symmetric.
                                                                  The proof of this fact is an immediate
                                  consequence of the preceding theorem and the results about existence,
                                   uniqueness and homomorphism property of  EH flows, as derived, e.g., in  \cite{GS},
                                    \cite{Par}. For more details
                                    regarding the dilation problem
                                    of symmetric q.d.s., the
                                    reader is referred to
                                    \cite{AF} and \cite{KM}.
                                 \ermrk

                      \section{A special class of  q.d.s. with unbounded generator on $\clb(h)$}
                It is not so easy to generalize the above results to the  case of q.d.s.
                   with unbounded generator due to a number of algebraic and analytic
                 difficulties. The first source of problem is the absence of any general structure
                  theorem like the theorem of Christensen-Evans which holds in the case of bounded generators. However,
                   for symmetric q.d.s. the situation is much better, because in this case a Christensen-Evans
                    type structure of the generator can be obtained
                     with an unbounded closable operator $R$. But the algebraic difficulty will remain since
                      we cannot ensure that $R^*\pi(x)R$ will be affiliated to the von Neumann algebra
                       $\cla$ (see \cite{GS0} for some examples where it is affiliated, although in many other
                        interesting  examples it will indeed not be affiliated!). Furthermore, if we embed
                        the range of $R$ into a factorizable Hilbert space of the form $h \ot k$ for some
                         $k$, then we do not have much control on the domain of $R_i^*$ (where we fix an orthonormal
                          basis $\{e_i\}$ of $k$ and write $R=\sum R_i \ot e_i$, with domain of each $R_i$ containing
                          domain of $R$), so that $R_i+R_i^*$ may not be densely defined. Thus, it is far from straightforward
                           how to look for a sufficient condition to obtain
                           a structure as in the bounded generator
                           case.
                           However, we shall present in what follows a special case in which such
                            a condition can be given.

                                Let $h$ be a separable Hilbert space,   $\cla=\clb(h)$, and $\tau$  be the usual
                                      trace of $\clb(h)$. Let us choose and fix some countable total subset of vectors $\cle$
                                       of $h$ and let $\clf_0$ denote the norm-dense $\ast$-subalgebra of
                                       $\clb_0(h)$(the set of
                                       compcat operators on $h$)
                                     generated algebraically by  all  finite-rank operators of the form
                                      $|u><v|$ for $u,v \in \cle
                                      $. Let $\clf=\clf_0 \oplus
                                      1$. we have the following :
                                           \bppsn
                                                                  Let   $(T_t)_{t \geq 0}$ be a conservative q.d.s. on $\clb(h)$
                                     such that the generator $\cll$ has $\clf$ in its domain and $\cll(1)=0,$ $\cll(\clf_0)
                                      \subseteq \clf_0$. Assume furthermore that $\tau(\cll(a)b)=
                                       \tau(a\cll(b))$ for $a,b \in \clf$.  Then there exist $\ast$-derivations $\al_j,  j=1,2,...$ defined
                                           on $\clf$ such that $\al_j=\al_j^\dagger, \forall j$ (where $\al_j^\dagger(x)=
                                            \al_j(x^*)^*)$, and
                                            $<\xi,\cll(y)\eta>=$ $\frac{1}{2}\sum_j <\xi,\al_j^2(y)\eta>$
                                             for any $y \in \clf$  and $\xi,\eta$ belonging to
                                              the domain consisting of finite linear combinations of
                                              vectors in $\cle$.
                                            \eppsn
                                          {\it Proof :}\\
                                                                             By standard arguments as in
                                        \cite{Sav}, \cite{Sa1}, we can construct
                                        a separable Hilbert space $\clk$, a $\ast$-homomorphism $\pi : \clb_0(h) \raro
                                        \clb(k)$, and a $\pi$-derivation $\delta: \clf \raro \clb(h,\clk)$ such that
                                         $\cll(x^*y)-\cll(x)^*y-x^*\cll(y)=\delta(x)^*\delta(y)
                                          \forall x,y \in \clf$. Now, since every representation of $\clb_0(h)$ is  unitarily equivalent to a
                                       direct sum of the trivial representation, we can choose $\clk$ to be of the form $h \ot k$ for some separable
                                        Hilbert space $k$ and $\pi(x)=\Sigma^*(x \ot 1_k)\Sigma$ for some unitary $\Sigma$.
                                         But then by replacing $\delta$ by $\Sigma \delta$, we can assume without loss of generality
                                          that $\pi(x)=(x \ot 1)$ and $\delta$ is an $(x \ot 1)$-derivation, i.e. $\delta(x)u=\sum_i
                                           \delta_i(x)u \ot e_i, x \in \clf, u \in h$, where $\{e_i \}$  is an orthonormal basis of $k$ and $\delta_i
                                            : \clf \raro \clb(h)$ is a derivation for each $i$. In fact, $\delta_i(\clf) \subseteq \clb_0(h),$
                                             because $\cll$ maps $\clf$ into $\clb_0(h)$ (as $\cll(1)=0$ and $\cll$ maps $\clf_0$
                                          into $\clf_0$) and $0 \leq \delta_i(x)^* \delta_i(x) \leq
                                               \cll(x^*x)-\cll(x^*)x-x^*\cll(x)$, which is a finite-rank operator. Hence $\delta_i(x)^*\delta_i(x)$
                                                is compact, i.e. $\delta_i(x)$ is compact. Now, let us consider the following derivations defined on $\clf$, given by,
                                                    $\al_{2j}=\frac{1}{2}(\delta_j+\delta_j^\dagger); \al_{2j-1}=\frac{i}{2}
                                                    (\delta_j^\dagger-\delta_j), j=1,2,...$.
                                                    Clearly $\al_j^\dagger=\al_j$. We claim that for
                                                     $x,y \in \clf$, $\sum_j \al_j(x)\al_j(y)
                                                    $ weakly converges and equals $\cll(xy)-\cll(x)y-x\cll(y)$. The proof
                                                     is more or less the same as done earlier in context of q.d.s. with
                                                     bounded generator. First we
                                                      verify using symmetry of $\cll$  that for nonnegative $y \in \clf_0$, selfadjoint $x \in \clf$,
                                                       $0 \leq \tau(y\sum_{j=1}^n \delta_j(x)\delta_j^\dagger(x))\leq
                                                                                                                      \tau(\sum_{j=1}^\infty \delta_j^\dagger(x)y\delta_j(x)) =  \tau(y(\cll(x^2)-\cll(x)x
                                                       -x\cll(x)))$, and from this (since $\clf_0$ is clearly dense in the set of trace-class
                                                        operators in trace-norm) conclude that
                                                         the weak convergence of $\sum_j \delta_j(x)\delta_j^\dagger(x)
                                                         $ holds for all self-adjoint $x \in \clf$, and hence for all $x \in \clf$. The claim is then proved by a usual
                                                                              polarization argument.  Moreover,
                                                          we also get that $\sum_j \delta_j(x)\delta_j^\dagger(y)=\sum_j \delta_j^\dagger(x)
                                                          \delta_j(y) \forall x,y \in \clf$. This follows by showing that
                                                      $\tau(\sum_j \delta_j(x)\delta_j^\dagger(y)z)=\tau(\sum_j \delta_j^\dagger(x)
                                                          \delta_j(y)z) $ for all $z \in \clf_0$. Then  an easy computation as in the earlier section
                                                           enables us to verify the claim.

                                                             We
                                                             now
                                                             choose
                                                             an
                                                             orthonormal basis $\xi_k,k =1,2,...$ of $h$ such that each of this basis vector
                                                              belongs to the span of the total  set $\cle$ mentioned in the statement of the proposition, and  we denote by $\cla_n$
                                                               the algebra generated by $|u><v|, u,v \in span\{\xi_1,...\xi_n\}$ (thus $\cla_n$
                                                                is isomorphic with the algebra of $n \times n$ matrices). Since $\cll(\clf_0)\subseteq \clf_0$, clearly
                                                                 for each $m$, there is some $n$ with the
                                                                 property that
                                                                  $\cll(\cla_m) \subseteq \cla_n$.  It is easy to see that for any $j$, $\al_j$
                                                                   also maps $\cla_m$ into $\cla_n$.
                                                                    Take any self-adjoint element $x \in \cla_m$ and any vector
                                                                    $\xi$ orthogonal to $\clv_n:=span\{\xi_1,...\xi_n \}$.
                                                                    Since $\delta_j(x)^*\delta_j(x) \leq \cll(x^2)-\cll(x)x-
                                                                   x\cll(x) \in \cla_n$, we have $\delta_j(x)\xi=0$, and similarly
                                                                    from the inequality $\delta_j(x)\delta_j(x)^* \leq \cll(x^2)-\cll(x)x-
                                                                   x\cll(x) $ it follows that $\delta_j(x)^*\xi=0$. Thus $\al_j(x)\xi
                                                                   =0$ for any $\xi \in \clv_n^\perp$, and as $\al_j(x)$ is self-adjoint,
                                                                    it follows that $\al_j(x)(\clv_n) \subseteq \clv_n,
                                                                     $, i.e. $\al_j(x) \in \cla_n$ for all selfadjoint $x \in \cla_m$,
                                                                      and hence the same holds for any $x \in \cla_m$.

                                                                     But being a derivation from a
                                                                     finite dimensional matrix-algebra into another, there must exist $H_{m,n} \in \cla_n$
                                                                      which implements $\al_j|_{\cla_m}$, and thus we must have that
                                                                      $\tau(\al_j(x))=0 \forall x \in \cla_m$.
                                       Now it is simple  to verify that for $x,y \in \cla_m$, $2\tau(x \cll(y))=-\sum_j
                                                              \tau(\al_j(x)\al_j(y))=\sum_j \tau(x \al_j^2(y)) $,
                                                               where in the last step of the above equality we have used the fact
                                                                that $\tau(\al_j(x\al_j(y)))=0$. This proves that on the algebraic direct sum
                                                                 of $\cla_m, m=1,2,...$ $\cll$ is indeed of the form $\frac{1}{2}\sum_j \al_j^2$,
                                                                  in the sense
                                                                  that the sum $\sum_j \frac{1}{2} <\xi, \al_j^2(y)\eta>$ converges to
                                                                   $<\xi,\cll(y)\eta>$ for every $y$ in the algebraic direct sum of
                                                                   $\cla_m,m=1,2,...$, and
                                                                    for $\xi, \eta$ belonging to the dense set consisting of finite linear combinations
                                                                     of
                                                                     $\xi_k,k=1,2,...$.\\

                                                             {\it
                                                             Example
                                                             :}\\
                                                                     Let us now give an example
                                                                     where the assumptions of the above
                                                                     Proposition
                                                                     are
                                                                     valid.
                                                                     Let
                                                                     $h=L^2(R^N)$,
                                                                     $\cld$
                                                                     be
                                                                     some
                                                                     countable
                                                                     dense
                                                                     subset
                                                                     of
                                                                     $h$ consisting of elements from $C_c^\infty(R^N)$ (the set of smooth functions with compact supports),
                                                                     and let $\cle=\bigcup_{n\geq 0}\bigcup_{\al : |\al |=n}
                                                                     \{ f^{(\al)} : f \in \cld \},$ where $\al$ denotes a multi-index, say $\al=(\al_1,...\al_N), \al_j \geq
                                                                     0,
                                                                     |\al|=\al_1+...\al_N$, and
                                                                      $f^{(\al)}:=(\frac{\partial}{\partial x_1})^{\al_1}...(\frac{\partial}{\partial x_N})^{\al_N}f$ and $f^{(0,...0)}=f$.
                                                                       Let $H_j,j=1,...,N$ denote the anti-selfadjoint operator $\frac{\partial}{\partial x_j}$ on $h$,
                                                                        and we consider the
                                                                        formal
                                                                        generator
                                                                        $\cll$
                                                                        given
                                                                        by
                                                                        $\cll=\frac{1}{2}
                                                                        \sum_{j=1}^N
                                                                        [
                                                                        H_j,[H_j,.]]$
                                                                       on
                                                                         the
                                                                         linear
                                                                         span
                                                                         of
                                                                         $\{|u><v|,
                                                                         u,v
                                                                         \in
                                                                         \cle
                                                                         \}$.
                                                                         By
                                                                         the
                                                                         results
                                                                         of
                                                                         \cite{Kat},
                                                                         it
                                                                         is
                                                                         indeed
                                                                         possible
                                                                         to
                                                                         obtain
                                                                         a
                                                                         conservative
                                                                         q.d.s.
                                                                         $T_t$
                                                                         on
                                                                         $\clb(h)$
                                                                         whose
                                                                         generator
                                                                         extends
                                                                         $\cll$,
                                                                         and
                                                                         it
                                                                         is
                                                                         easy
                                                                         to
                                                                         see
                                                                         that
                                                                         the
                                                                         hypotheses
                                                                         of
                                                                         the
                                                                         above
                                                                         Proposition
                                                                         are
                                                                         satisfied.

                                                                We conclude with a few remarks on the possibility of obtaining a similar result for q.d.s. with
                                                                 unbounded generators on von Neumann or $C^*$ algebras other than the special cases considered here.
                                                                  There is a serious problem to even conceive of an appropriate generalization. In \cite{unbdd},
                                                                    an interesting class of symmetric q.d.s. on general $C^*$
                                                                    or von Neumann algebras with an additional assumption of covariance w.r.t. the action of
                                                                     a (possibly noncommutative and noncompact) Lie group has been studied in detail,
                                                                     and    results
                                                                      regarding the structure of the generator as well as the existence and homomorphism property of
                                                                       EH dilation have been obtained.  For technical reasons  the authors in \cite{unbdd}
                                                                        had to work with
                                                                        the canonical embedding of the algebra in the $L^2$-space of the given trace, and
                                                                         the methods would not apply if   any other embedding was considered.
                                                                          Thus, if we want to
                                                                          imitate the techniques used in this present article to obtain structure theorems for
                                                                           q.d.s. with unbounded generators, it is in principle possible to give some sufficient
                                                                            conditions on the generator in order to be able to write it as a series of squares of
                                                                             derivations only in the case where either the algebra or its commutant is
                                                                              abelian. But remembering that to apply the set-up of \cite{unbdd},
                                                                               we must take the commutant w.r.t. the embedding in the $L^2$-space
                                                                                of the trace, which is by the Tomita-Takesaki theory anti-isomorphic
                                                                                 to the algebra itself, it is easily seen that we are confined to the case of
                                                                                  abelian algebras only ! Thus, the question how to adapt the technique of section 2
                                                                                   to the case of unbounded generators for general algebras remains open; either one has
                                                                                    to improve or modify the results of section 2 by weakening the assumption of commutativity
                                                                                     of the algebra or its commutant, or to improve the techniques of \cite{unbdd} to accommodate
                                                                                      more general situations.

{\bf Acknowledgement :\\}
Debashish Goswami would like to thank Prof.
    S. Albeverio for inviting him to I.A.M. (Bonn) and Av Humboldt Foundation
     for the Research Fellowship for the period November 2000 to October 2001.

Institut f{\"u}r Angewandte Mathematik, Universit{\"a}t Bonn, Wegelerstr. 6,
 D-53115, Bonn, Germany.\\
email: albeverio@wiener.iam.uni-bonn.de,\\
goswamid@wiener.iam.uni-bonn.de

\end{document}